\documentclass[aps,prd,twocolumn,floatfix,superscriptaddress,nofootinbib]{revtex4-2}

\usepackage{graphicx}
\usepackage{amsmath}
\usepackage{amssymb}
\usepackage{xcolor}
\usepackage[colorlinks,citecolor=blue,linkcolor=blue,urlcolor=blue,anchorcolor=blue]{hyperref}
\allowdisplaybreaks
\newcommand\dx{{\rm d}}
\newcommand\p{\partial}
\newcommand\etal{{\it et~al.}}

\begin{document}

\title{Early dark energy in $k$-essence}
\author{S. X. Tian}
\email[]{tshuxun@bnu.edu.cn}
\affiliation{Department of Astronomy, Beijing Normal University, 100875 Beijing, China}
\author{Zong-Hong Zhu}
\email[]{zhuzh@bnu.edu.cn}
\affiliation{Department of Astronomy, Beijing Normal University, 100875 Beijing, China}
\affiliation{School of Physics and Technology, Wuhan University, 430072 Wuhan, China}
\date{\today}
\begin{abstract}
  Early dark energy (EDE) that becomes subdominant around the epoch of matter-radiation equality can be used to ease the Hubble tension. However, there is a theoretical problem that why the energy scale of EDE is in coincidence with that of matter-radiation equality when their physics are completely unrelated. Sakstein and Trodden [\href{https://doi.org/10.1103/PhysRevLett.124.161301}{Phys. Rev. Lett. {\bf 124}, 161301 (2020)}] proposed a mechanism to solve this coincidence problem with $\mathcal{O}({\rm eV})$-mass neutrino. In this paper, in order to solve the coincidence problem, we propose a new scenario for EDE, in which the onset and ending of EDE are triggered by the radiation-matter transition. The specific example we study is a $k$-essence model. The cosmic evolution equations can be recast into a two-dimensional dynamical system and its main properties are analyzed. Our results suggest that $k$-essence seems unable to realize the new scenario for EDE. However, an EDE model with different scenario is realized in $k$-essence. In this model, the ending of EDE can be triggered by the radiation-matter transition while the onset depends on the initial conditions of the scalar field. Therefore, the obtained model can only be used to solve half of the coincidence problem. The full resolution in the framework of our initial proposed scenario is worthy of more research.
\end{abstract}
\pacs{}
\maketitle

\section{Introduction}\label{sec:01}
Today we are entering the era of precise cosmology. Many observations, such as type Ia supernovae (SNe Ia \cite{Scolnic2018.ApJ.859.101}), baryon acoustic oscillations \cite{Ata2017.MNRAS.473.4773}, cosmic microwave background (CMB \cite{Aghanim2018.arXiv.1807.06209}), gravitational lensing systems \cite{Birrer2019.MNRAS.484.4726}, gravitational waves \cite{Abbott2017.Nature.551.85} and so on, can give tight constraints on the cosmological parameters. However, a discrepancy between high and low redshift measurements of the Hubble constant $H_0$ appeared with the improvement of precision. The \textit{Planck} CMB data, together with the $\Lambda$CDM model, predict $H_0=67.4\pm0.6\ {\rm km}/{\rm s}/{\rm Mpc}$ \cite{Aghanim2018.arXiv.1807.06209}, while the local Cepheid-calibrated SNe Ia observations give $H_0=74.03\pm1.42\ {\rm km}/{\rm s}/{\rm Mpc}$ \cite{Riess2019.ApJ.876.85}. Statistically, there is an inconsistency of approximately $4\sigma$ level between these two measurements. This is called the Hubble tension and is one of the most important issue in modern cosmology \cite{DiValentino2017.NAA.1.569}.

If, as suggested by the existing analysis \cite{Spergel2015.PRD.91.023518,Addison2016.ApJ.818.132,Rigault2015.ApJ.802.20,Jones2018.ApJ.867.108}, systematic errors in observations are not responsible for the inconsistency, then the Hubble tension may indicate new physics beyond the $\Lambda$CDM model. There are many attempts to address the Hubble tension rely on the possible new physics, such as Planck-scale loop quantum cosmology \cite{Ashtekar2020.PRL.125.051302}, primordial magnetic field \cite{Jedamzik2020.PRL.125.181302}, early dark energy (EDE \cite{Poulin2019.PRL.122.221301,Braglia2020.PRD.102.083513,Niedermann2020.PRD.102.063527,Smith2020.PRD.101.063523,Zumalacarregui2020.PRD.102.023523,
Braglia2020.arXiv.2011.12934,Chudaykin2020.arXiv.2011.04682}), low redshift resolutions \cite{Vagnozzi2020.PRD.102.023518,Wang2020.PRD.102.063530,Yang2020.PRD.102.063503,DeFelice2020.arXiv.2009.08718,Yao2020.arXiv.2011.09160} and so on \cite{Ye2020.PRD.101.083507,Ye2020.PRD.102.083523,Haslbauer2020.arXiv.2009.11292}. Among these, EDE has attracted widespread attention. A scalar field with up to approximately $10\%$ relative energy density at a critical redshift $z_c\sim3500$ seems able to bring the high and low redshift measurements of $H_0$ into agreement \cite{Smith2020.PRD.101.063523}. The underlying mechanism of the EDE resolution is that it can reduce the cosmic sound horizon at last scattering and thus increase the measured value of $H_0$ form CMB data. However, controversy about EDE's ability to ease the Hubble tension exists in the literature \cite{Hill2020.PRD.102.043507,Ivanov2020.PRD.102.103502,Knox2020.PRD.101.043533,Jedamzik2020.2010.04158,Smith2020.2009.10740}. In this paper, we ignore this controversy and assume that EDE can indeed ease the Hubble tension. The topic of this paper is to discuss the theoretical naturalness of EDE. More precisely, our discussion focus on the coincidence problem about EDE, which was first recognized in Ref. \cite{Sakstein2020.PRL.124.161301}.

Historically, there has been some works discussing EDE long before the confirmation of the Hubble tension \cite{Doran2001.ApJ.559.501,Wetterich2004.PLB.594.17,Doran2006.JCAP.06.026,Pettorino2013.PRD.87.083009,Feng2011.RAA.11.751}. These works generally described EDE with parameterized relative energy density $\Omega_{\rm EDE}$ or equation of state $w_{\rm EDE}$, and the time when EDE appears was not determined based on any theoretical or observational reasons. Hubble tension determined the emergence time of EDE to be near the matter-radiation equality \cite{Poulin2019.PRL.122.221301}. However, a coincidence problem related to EDE also emerged simultaneously \cite{Sakstein2020.PRL.124.161301}. Intuitively, the physics of the matter-radiation equality is completely irrelevant to that of EDE, but why do they appear at the same time? Technically, one can specify the energy scale\footnote{Note that the emergence time of EDE in cosmic history is specified by its energy scale.} of EDE by introducing the cosmic expansion rate $H(z_c)$ into the parameterized $\Omega_{\rm EDE}$ or $w_{\rm EDE}$ \cite{Poulin2019.PRL.122.221301}, or even a first principle Lagrangian \cite{Braglia2020.PRD.102.083513,Smith2020.PRD.101.063523}, which is proposed after the confirmation of the Hubble tension. This can give a working EDE model, but does not help solve the related coincidence problem. Is it possible to construct EDE around the matter-radiation equality without introducing $H(z_c)$? Refs. \cite{Sakstein2020.PRL.124.161301,Gogoi2020.arXiv.2005.11889,Gonzalez2020.arXiv.2011.09895} did it. They specified the energy scale of EDE through a nonminimal coupling between $\mathcal{O}({\rm eV})$-mass neutrino and a scalar field. In this model, EDE appears when neutrinos become nonrelativistic, i.e., at the energy scale of $\mathcal{O}({\rm eV})$, which is exactly around the matter-radiation equality. Therefore, this neutrino-assisted model can naturally solve the coincidence problem about EDE.

In this paper, we try a different approach to solve the coincidence problem about EDE. Look back the cosmic history, the epoch of matter-radiation equality is special. At that time, the Universe changed from radiation-dominated to matter-dominated. We want use the radiation-matter transition to trigger the onset and ending of EDE. If so, the energy scale of EDE must be in coincidence with the energy scale of matter-radiation equality. Furthermore, we require the model only introduces Planck scale parameters and dimensionless parameters of order unity. This requirement has been imposed to construct dark energy models that can solve the coincidence problem related to the cosmic late-time acceleration \cite{Tian2020.PRD.101.063531,Tian2020.PRD.102.063509}. In this scenario, no explicit energy scale below Planck values is required, which is an essential difference between our model and the existing models, e.g., EDE models we mentioned in the previous two paragraphs. To achieve the above scenario, one general $k$-essence model is discussed in Sec. \ref{sec:02} and one specific case is analyzed in Sec. \ref{sec:03}.

\section{One general model}\label{sec:02}
\begin{figure}[t]
  \centering
  \includegraphics[width=0.9\linewidth]{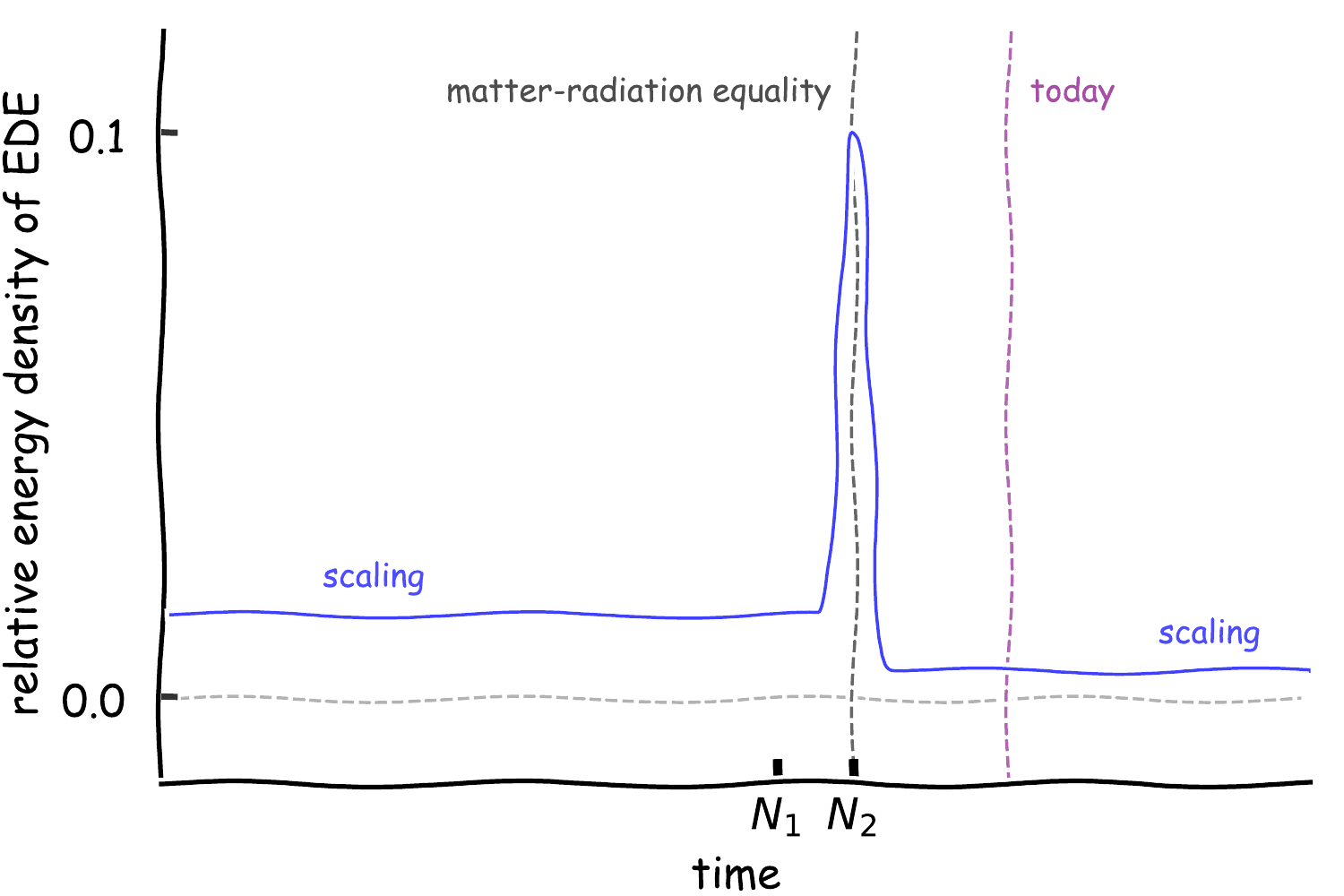}
  \caption{Illustration of our initial idea about the EDE (see the main text for details).}
  \label{fig:01}
\end{figure}

A beautiful theoretical model should be as simple as possible. As the first step, we only consider the minimally coupled single scalar field in this paper. The model should  belong to $k$-essence \cite{Armendariz-Picon1999.PLB.458.209,Armendariz-Picon2000.PRL.85.4438,Armendariz-Picon2001.PRD.63.103510,Li2011.CTP.56.525}. We expect that the EDE relative energy density $\Omega_{\rm EDE}$ has a peak near the matter-radiation equality, and keeps small enough in the radiation or matter-dominated Universe. For a minimally coupled EDE, its energy conservation equation can be formally written as $\dot{\rho}_{\rm EDE}+3(1+w_{\rm EDE})H\rho_{\rm EDE}=0$. Introducing $N=\ln(a/a_0)$, this equation gives
\begin{equation}\label{eq:01}
  \frac{\rho_{\rm EDE}(N_2)}{\rho_{\rm EDE}(N_1)}=\exp\left[-3\int_{N_1}^{N_2}[1+w_{\rm EDE}(\tilde{N})]\dx\tilde{N}\right],
\end{equation}
where $N_2$ represents the time when EDE appears, and $N_1$ is slightly earlier than $N_2$ (see Fig. \ref{fig:01} for an illustration).
Even if $w_{\rm EDE}$ can be less than $-1$, the right side of Eq. (\ref{eq:01}) will not be a very large number. Therefore, if observation expects $\rho_{\rm EDE}(N_2)$ to be considerable compared with normal matters \cite{Poulin2019.PRL.122.221301}, then $\rho_{\rm EDE}(N_1)$ cannot be extremely small compared with normal matters. One way to preserve a non-vanishing $\Omega_{\rm EDE}(N_1)$ in the radiation-dominated Universe is the scaling solution \cite{Copeland1998.PRD.57.4686}. The realistic EDE may evolve as a scaling solution with small\footnote{We require $\Omega_{\rm EDE}$ to be small in the radiation-dominated era due to the primordial nucleosynthesis constraints \cite{Copeland1998.PRD.57.4686}.} $\Omega_{\rm EDE}$ in the radiation-dominated era. Similarly, EDE may also evolve as a scaling solution in the matter-dominated era. Ref. \cite{Piazza2004.JCAP.07.004} proved Eq. (\ref{eq:03}) is necessary for the existence of the scaling solution in $k$-essence. Therefore, in order to preserve the scaling solution, we adopt the action
\begin{equation}\label{eq:02}
  S=\int\dx^4x\sqrt{-g}\left[\frac{R}{2\kappa}-\frac{\mathcal{L}}{\kappa}\right]+S_{\rm m},
\end{equation}
where $\kappa=8\pi G/c^4$ and
\begin{equation}\label{eq:03}
  \mathcal{L}=[1+f(y)]X+[1+g(y)]V_{\rm exp},
\end{equation}
where $X=\frac{1}{2}g^{\mu\nu}\nabla_\mu\phi\nabla_\nu\phi$, $V_{\rm exp}=V_0\exp(-\lambda\phi)$, $V_0$ and $\lambda$ are constants, $y=X/V_{\rm exp}$, $f(y)$ and $g(y)$ are arbitrary functions. Performing the transformation
\begin{equation}
  g(y)=-f(y)y+h(y),
\end{equation}
the above Lagrangian can be rewritten as
\begin{equation}\label{eq:03b}
  \mathcal{L}=X+h(y)V_{\rm exp}.
\end{equation}
This transformation eliminates one of the arbitrary functions in the Lagrangian, as well as in the corresponding cosmic evolution equations. In our conventions, $\lambda$ is dimensionless and $[V_0]={\rm length}^{-2}$. As the following results do not explicitly depend on the value of $V_0$, we can assume $V_0=\mathcal{O}(l_{\rm P}^{-2})$, where $l_{\rm P}$ is the Planck length. Similar property exists in the exponential quintessence model \cite{Copeland1998.PRD.57.4686} and an oscillating dark energy model \cite{Tian2020.PRD.101.063531,Tian2020.PRD.102.063509}. For $S_{\rm m}$, we know its variation $\delta S_{\rm m}=-\frac{1}{2}\int\dx^4x\sqrt{-g}T_{\mu\nu}\delta g^{\mu\nu}$ and $T_{\mu\nu}=\left(\rho_{\rm m}+p_{\rm m}/c^2\right)u_\mu u_\nu+p_{\rm m}g_{\mu\nu}$. The equation of state of normal matters is defined as $w_{\rm m}\equiv p_{\rm m}/(\rho_{\rm m}c^2)$. For the realistic Universe, normal matters include radiation $(\rho_{\rm rad}\propto e^{-4N})$ and pressureless fluid $(\rho_{\rm mat}\propto e^{-3N})$. Therefore, the $w_{\rm m}$ can be written as \cite{Tian2020.PRD.102.063509}
\begin{equation}\label{eq:wm}
  w_{\rm m}(N)=\frac{\rho_{\rm rad}c^2/3+0}{(\rho_{\rm rad}+\rho_{\rm mat})c^2}=\frac{1/3}{1+e^{N-N_{\rm eq}}},
\end{equation}
where $N_{\rm eq}$ corresponds to the matter-radiation equality. Lagrangian similar to Eq. (\ref{eq:03b}) has been used to construct dark energy for the late-time Universe \cite{Piazza2004.JCAP.07.004,Tamanini2014.PRD.89.083521,Bahamonde2018.PREP.775.1}. Setting $h(y)=0$, Eq. (\ref{eq:03b}) gives the classical exponential quintessence model \cite{Copeland1998.PRD.57.4686}. Nonzero $h(y)$ can make the motion of the scalar field more complicated when $w_{\rm m}$ goes from $1/3$(radiation) to 0(matter). We expect that the radiation-matter transition with nonzero $h(y)$ can induce the desired EDE. The above scenario is summarized in Fig. \ref{fig:01}. Note that this is just our initial idea about the EDE, and the final result obtained in this paper is slightly different from it.

Variation of Eq. (\ref{eq:02}) with respect to the metric gives the gravitational field equations
\begin{equation}\label{eq:04}
  G_{\mu\nu}=\kappa T_{\mu\nu}+\Phi_{\mu\nu},
\end{equation}
where $\Phi_{\mu\nu}=\mathcal{L}_X\p_\mu\phi\p_\nu\phi-\mathcal{L}g_{\mu\nu}$ and $\mathcal{L}_X\equiv\p\mathcal{L}/\p X$. The equation of motion of the scalar field is
\begin{equation}\label{eq:05}
  (\p_\mu\mathcal{L}_X)\p^\mu\phi+\mathcal{L}_X\Box\phi-\mathcal{L}_\phi=0,
\end{equation}
where $\mathcal{L}_\phi\equiv\p\mathcal{L}/\p\phi$. In this paper, we assume the Universe is described by the flat Friedmann-Lema\^{i}tre-Robertson-Walker metric
\begin{equation}\label{eq:06}
  \dx s^2=-c^2\dx t^2+a^2(\dx x^2+\dx y^2+\dx z^2),
\end{equation}
where $a=a(t)$. Substituting Eq. (\ref{eq:06}) and $\phi=\phi(t)$ into Eqs. (\ref{eq:04}) and (\ref{eq:05}), we obtain the cosmic evolution equations
\begin{gather}
  H^2=\frac{\kappa c^4}{3}\rho_{\rm tot}, \label{eq:07}\\
  \frac{\ddot{a}}{a}=-\frac{\kappa c^4}{6}\left(\rho_{\rm tot}+\frac{3p_{\rm tot}}{c^2}\right), \label{eq:08}\\
  \ddot{\phi}(\mathcal{L}_X+2X\mathcal{L}_{XX})+3H\dot{\phi}\mathcal{L}_X\quad\nonumber\\
  \quad\qquad\qquad+c^2(\mathcal{L}_\phi-2X\mathcal{L}_{X\phi})=0, \label{eq:09}
\end{gather}
where $\rho_{\rm tot}=\rho_{\rm m}+\rho_\phi$, $p_{\rm tot}=p_{\rm m}+p_\phi$, $\rho_\phi=(\mathcal{L}-2X\mathcal{L}_X)/(\kappa c^2)$, $p_\phi=-\mathcal{L}/\kappa$, $\mathcal{L}_{XX}\equiv\p^2\mathcal{L}/\p X^2$ and $\mathcal{L}_{X\phi}\equiv\p^2\mathcal{L}/(\p\phi\p X)$. Equation (\ref{eq:09}) can be derived from Eqs. (\ref{eq:07}) and (\ref{eq:08}). Following Refs. \cite{Copeland1998.PRD.57.4686,Tian2020.PRD.101.063531}, we introduce
\begin{equation}
  x_1\equiv\frac{\dot{\phi}}{\sqrt{6}H},\quad
  x_2\equiv\frac{c\sqrt{V_{\rm exp}}}{\sqrt{3}H}.
\end{equation}
This definition gives $x_2>0$. Then the cosmic evolution equations can be written as
\begin{align}
  \frac{\dx x_1}{\dx N}&=\frac{-3x_1(h'+1)+\sqrt{3/2}\lambda x_2^2(2y^2h''-yh'+h+1)}{2yh''+h'+1}\nonumber\\
  &\quad-\frac{3}{2}x_1L, \label{eq:14}\\
  \frac{\dx x_2}{\dx N}&=-\frac{\sqrt{6}}{2}\lambda x_1x_2-\frac{3}{2}x_2L, \label{eq:15}
\end{align}
where $'\equiv\dx/\dx y$, $y=-x_1^2/x_2^2$ and $L=2w_{\rm m}x_1^2h'+(1+w_{\rm m})(x_2^2h+x_2^2-1)+(w_{\rm m}-1)x_1^2$. The EDE relative energy density is defined as $\Omega_\phi\equiv8\pi G\rho_\phi/(3H^2)$ and its explicit expression is
\begin{equation}\label{eq:16}
  \Omega_\phi=x_2^2(-2yh'+h-y+1).
\end{equation}
Note that the above results are valid for both constant and time-varying $w_{\rm m}$.

The scenario depicted in Fig. \ref{fig:01} strongly depends on the scaling solution. A scaling solution corresponds to a critical point of the above dynamic system with $0<\Omega_\phi<1$. Solving $\dx x_1/\dx N=0$ and $\dx x_2/\dx N=0$, we obtain one critical point
\begin{align}
  x_{1,{\rm c}}&=\frac{\sqrt{6}(1+w_{\rm m})}{2\lambda},\\
  x_{2,{\rm c}}&=\sqrt{\frac{3(1+w_{\rm m})(1-w_{\rm m}-2w_{\rm m}h')}{2\lambda^2(h+1)}}, \label{eq:18}
\end{align}
which corresponds to the scaling solution, and
\begin{equation}\label{eq:19}
  \Omega_{\phi,{\rm c}}=\frac{3(1+w_{\rm m})(1+h')}{\lambda^2}.
\end{equation}
Note that Eq. (\ref{eq:18}) is an implicit equation as $h=h(-x_{1,{\rm c}}^2/x_{2,{\rm c}}^2)$. Without the expression of $h(y)$, one cannot obtain the explicit coordinates, nor analyze its stability and existence conditions.

Our original motivation was to construct a model to realize the scenario described in Fig. \ref{fig:01}. However, we are failed to achieve this. We tested several $h(y)$: $h(y)=\exp(\alpha y)$, $\cos(\alpha y)$ and $-\alpha(y+\beta)^n$ for $n=2,3,4$. The functions $\exp(\alpha y)$ and $-\alpha(y+\beta)^3$ are monotonic with respect to $y$, while other functions are not. Contrary to Fig. \ref{fig:01}, there is no peak between the two scaling solutions for all the tests. $\Omega_\phi$ evolves almost\footnote{Generally, there is a small oscillation around $\Omega_\phi=\Omega_{\phi,{\rm c},0}$. This is understandable because the corresponding critical point is an attracting spiral (see the exponential quintessence \cite{Copeland1998.PRD.57.4686} for an example).} monotonously from $\Omega_{\phi,{\rm c},1/3}$ to $\Omega_{\phi,{\rm c},0}$, where the subscripts $1/3$ and $0$ represent the value of $w_{\rm m}$. But, in these tests, we also found some interesting results. For some specific $h(y)$, the ratio $r\equiv\Omega_{\phi,{\rm c},0}/\Omega_{\phi,{\rm c},1/3}$ can be extremely small. Note that $r=3/4=\mathcal{O}(1)$ for the exponential quintessence model (see Eq. (\ref{eq:19}) with $h=0$ or Ref. \cite{Copeland1998.PRD.57.4686}). With small enough $r$, we propose a new scenario to realize the EDE in Sec. \ref{sec:03}. The relevant coincidence problem is also discussed.

\section{Dynamics of a specific model}\label{sec:03}
In this section, we assume $h(y)=-\alpha(y+\beta)^2$, where $\alpha$ and $\beta$ are constants. There is no essential difference for the cosmic evolutions between this and other $h(y)$ functions. The function $h(y)=-\alpha(y+\beta)^2$ can be regarded as a representative of the functions we tested. Numerical results show that the expected model lies in the parameter space of $\lambda\gg1$, $\alpha\geq0$ and $\beta\geq0$. Therefore, we only discuss this parameter space for simplicity. Substituting the expression of $h(y)$ into Eq. (\ref{eq:18}), we can find the explicit coordinates of the critical point. For $w_{\rm m}=1/3$, we obtain
\begin{align}
  x_{1,{\rm c},1/3}=\frac{2\sqrt{6}}{3\lambda},\quad
  x_{2,{\rm c},1/3}=\frac{2}{\sqrt{3}\lambda}\sqrt{\frac{1-2\alpha\beta}{1-\alpha\beta^2}},
\end{align}
and
\begin{equation}\label{eq:21}
  \Omega_{\phi,{\rm c},1/3}=\frac{4+16\alpha(1-\beta)}{\lambda^2(1-2\alpha\beta)}.
\end{equation}
The existence of the above critical point requires $0<\Omega_{\phi,{\rm c},1/3}<1$ and $x_2$ is a positive real number. Linear stability theory \cite{Bahamonde2018.PREP.775.1} is suitable for analyzing the stability of this critical point. It is stable if the real part of the corresponding eigenvalues are all negative. The result is that the above critical point exists and is stable when
\begin{equation}\label{eq:22}
  \left\{ \begin{array}{ll}
  0\leq\beta<1/\sqrt{\alpha} & \textrm{if $0<\alpha<1/4$ \& $\lambda\gg1$},\\
  0\leq\beta<\frac{\lambda^2-16\alpha-4}{2\alpha(\lambda^2-8)} & \textrm{if $\frac{1}{4}<\alpha<\frac{\lambda^2-4}{16}$ \& $\lambda\gg1$}.
  \end{array} \right.
\end{equation}
For $w_{\rm m}=0$, we obtain
\begin{equation}\label{eq:23}
  \Omega_{\phi,{\rm c},0}=\frac{3}{\lambda^2}\sqrt{1+4\alpha(1-\beta)},
\end{equation}
and the corresponding critical point exists and is stable when
\begin{equation}\label{eq:24}
  \left\{ \begin{array}{ll}
  0\leq\beta<1/\sqrt{\alpha} & \textrm{if $0<\alpha<1/4$ \& $\lambda\gg1$},\\
  0\leq\beta<1+1/(4\alpha) & \textrm{if $1/4<\alpha$ \& $\lambda\gg1$}.
  \end{array} \right.
\end{equation}
Parameter space Eq. (\ref{eq:24}) always contains parameter space Eq. (\ref{eq:22}). Therefore, Eq. (\ref{eq:22}) is sufficient to ensure the existence of the above two stable critical points.

\begin{figure}[t]
  \centering
  \includegraphics[width=0.9\linewidth]{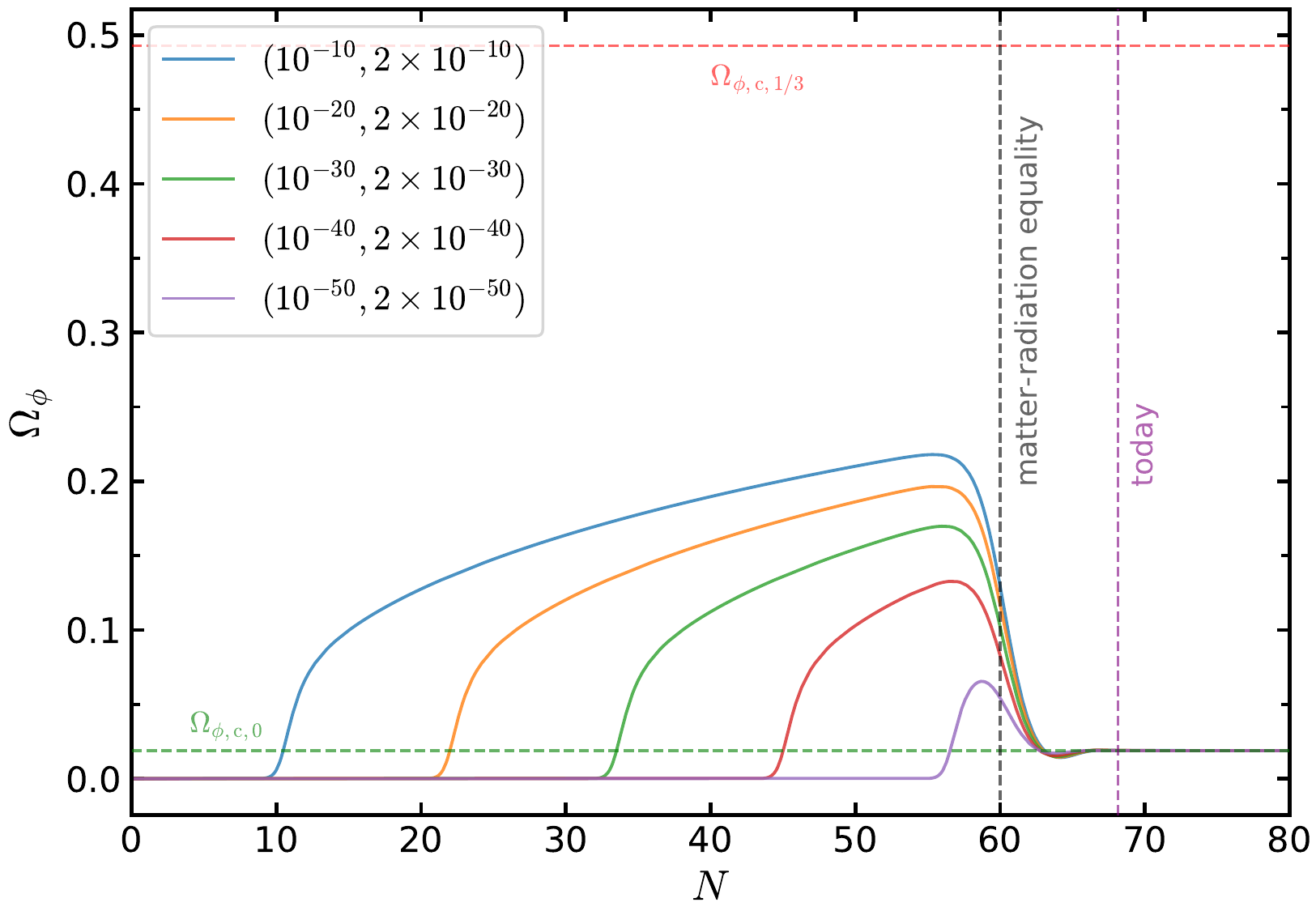}
  \caption{Evolution of the EDE relative energy density $\Omega_\phi$. The model parameters are $\lambda=25$, $\alpha=4$ and $\beta=0.1$. The initial conditions of $(x_1,x_2)$ are given in the legend. The equation of state $w_{\rm m}$ is given by Eq. (\ref{eq:wm}) with $N_{\rm eq}=60$. This scenario solves half of the coincidence problem we discussed in Sec. \ref{sec:01}.}
  \label{fig:02}
\end{figure}

EDE means a peak of $\Omega_\phi$ near matter-radiation equality. Numerical result shows that there is no peak between the two scaling solutions for the model with $h(y)=-\alpha(y+\beta)^2$. Here we propose a new scenario to realize the EDE in the framework of $h(y)=-\alpha(y+\beta)^2$. Inspired by Ref. \cite{Agrawal2018.PLB.784.271}, we can set the initial conditions of $(x_1,x_2)$ to be very close to the repulsive fixed point $(0,0)$ to suppress $\Omega_\phi$ in the far past. Meanwhile, if we set $\Omega_{\phi,{\rm c},1/3}\approx10\%$ and $\Omega_{\phi,{\rm c},0}\ll\Omega_{\phi,{\rm c},1/3}$ with suitable parameters, then the scalar field will evolve as the expected EDE. In Fig. \ref{fig:02}, we plot the evolution of $\Omega_\phi$ for several cases. The peak appears near the matter-radiation equality as we expected. The relative energy density $\Omega_\phi$ decreases rapidly after the matter-radiation equality. This solves half of the coincidence problem discussed in Sec. \ref{sec:01} that why the EDE decreases after the matter-radiation equality. Note that we solve this problem without specifying any energy scale. However, the remaining half has not been resolved. Our model cannot naturally explain the onset of EDE, which depends on the initial conditions of $(x_1,x_2)$, i.e., the initial energy density, in our model. In addition, one thing is worth mentioning here. Fig. \ref{fig:02} shows that the attractive ability of the stable critical point with $w_{\rm m}=1/3$ is weak. Therefore, if we expect the peak value of $\Omega_\phi$ to be approximately equal to $10\%$, then we may need to set $\Omega_{\phi,{\rm c},1/3}$ slightly larger than $10\%$.

\section{Conclusions}
In this paper we have tried a new mechanism to solve the coincidence problem about EDE, which is first discussed in Ref. \cite{Sakstein2020.PRL.124.161301}. Our idealized mechanism is to use the radiation-matter transition to trigger the onset and ending of EDE, so as to solve the related coincidence problem. The desired model is illustrated in Fig. \ref{fig:01}. No energy scale needs to be specified in the idealized mechanism. As a comparison, Ref. \cite{Sakstein2020.PRL.124.161301} uses neutrino with $\mathcal{O}({\rm eV})$ rest mass to specify the energy scale of EDE. We tried to use the $k$-essence described by Eq. (\ref{eq:03b}) to achieve the above mechanism. However, we were not completely successful in constructing the model. The problem is that, for the model we considered, if the scalar field evolves as scaling solution in both the radiation and matter-dominated era, then no peak of $\Omega_\phi$ appears around matter-radiation equality. One good news is that the radio $r\equiv\Omega_{\phi,{\rm c},0}/\Omega_{\phi,{\rm c},1/3}$ can be extremely small for some parameter settings. This property, together with the method that using initial conditions to suppress $\Omega_\phi$ in the far past \cite{Agrawal2018.PLB.784.271}, inspired us to propose a new framework to achieve EDE. Several illustrative examples are plotted in Fig. \ref{fig:02}. In this scenario, the onset of EDE depends on the initial conditions, which means the coincidence problem related the onset of EDE is not solved, while the ending is always around the matter-radiation equality, which means the coincidence problem related the ending of EDE can be solved naturally.

Our initial idea about EDE that using the radiation-matter transition to trigger its onset and ending is worthy of further exploration. The $k$-essence is a general minimally coupled model and it seems cannot realize our original idea. Nonminimal coupling may be required. For example, the Lagrangian
\begin{equation}
  \mathcal{L}=R+\underbrace{\alpha w_{\rm m}(1/3-w_{\rm m})R}_{\textrm{effective EDE}}
\end{equation}
may be able to realize our idea, where $\alpha=\mathcal{O}(1)$ is a dimensionless constant. In this model, the effective EDE may vanish in both radiation and matter-dominated era, but appear during the radiation-matter transition. The details will be carried out in a separate publication.

\section*{Acknowledgements}
This work was supported by the National Natural Science Foundation of China under Grants Nos. 11633001, 11920101003 and 12021003, the Strategic Priority Research Program of the Chinese Academy of Sciences, Grant No. XDB23000000 and the Interdiscipline Research Funds of Beijing Normal University. S.X.T. was supported by the Initiative Postdocs Supporting Program under Grant No. BX20200065.

%\bibliographystyle{apsrev4-2-2} % no title
%\bibliography{bibfile}
%apsrev4-2.bst 2019-01-14 (MD) hand-edited version of apsrev4-1.bst
%Control: key (0)
%Control: author (72) initials jnrlst
%Control: editor formatted (1) identically to author
%Control: production of article title (-1) disabled
%Control: page (0) single
%Control: year (1) truncated
%Control: production of eprint (0) enabled
%

\end{document}